\documentclass[twocolumn,showpacs,preprintnumbers,amsmath,amssymb]{revtex4}

\usepackage{graphicx}% Include figure files
\usepackage{dcolumn}% Align table columns on decimal point
\usepackage{bm}% bold math

\begin{document}

\title{Electronic Raman response in electron-doped cuprate
superconductors}

\author{Zhihao Geng and Shiping Feng}

\affiliation{Department of
Physics, Beijing Normal University, Beijing 100875, China}

%\date{\today}

\begin{abstract}
The electronic Raman response in the electron-doped cuprate
superconductors is studied based on the $t$-$t'$-$J$ model. It is
shown that although the domelike shape of the doping dependent peak
energy in the $B_{2g}$ symmetry is a common feature for both
electron-doped and hole-doped cuprate superconductors, there are
pronounced deviations from a cubic response in the $B_{1g}$ channel
and a linear response in the $B_{2g}$ channel for the electron-doped
case in the low energies. It is also shown that these pronounced
deviations are mainly caused by a nonmonotonic d-wave gap in the
electron-doped cuprate superconductors.
\end{abstract}


\maketitle

The parent compounds of cuprate superconductors are believed to
belong to a class of materials known as Mott insulators with an
antiferromagnetic (AF) long-range order (AFLRO), where a single
common feature is the presence of the CuO$_{2}$ plane
\cite{bednorz86,tokura89}. As the CuO$_{2}$ planes are doped with
charge carriers, holes or electrons, the AF phase subsides and
superconductivity emerges \cite{bednorz86,tokura89}. It has been
found that only an approximate symmetry in the phase diagram exists
about the zero doping line between the hole-doped ($p$-doped) and
electron-doped ($n$-doped) cuprate superconductors, and the
significantly different behavior of the $p$-doped and $n$-doped
cuprate superconductors is observed \cite{damascelli03}, reflecting
the electron-hole asymmetry. For the $p$-doped cuprate
superconductors, AFLRO is reduced dramatically with doping, and
vanished around the doping $\delta\sim 0.05$, then superconductivity
emerges over a wide range of the hole doping concentration $\delta$,
around the optimal doping $\delta\sim 0.15$
\cite{bednorz86,damascelli03}, where the pair-breaking peaks
observed from the Raman scattering measurements have been extracted
in different polarization configurations
\cite{chen98,guyard08,venturini02,guyard08b,tacon06}, and the
results show that the electronic Raman response (ERR) depends
linearly on energy in the $B_{2g}$ channel in the low energy limit,
and depends cubically on energy in the $B_{1g}$ channel. However,
AFLRO survives until superconductivity appears over a narrow range
of the electron doping concentration $\delta$, around the optimal
doping $\delta\sim 0.15$ in the $n$-doped counterparts
\cite{tokura89,damascelli03}, where although the low-energy
behaviors of the $B_{1g}$ and $B_{2g}$ channels approach very
specific power laws consistent with the presence of lines nodes in
the superconducting (SC) gap, the pair-breaking peak energy values
in the $B_{1g}$ and $B_{2g}$ channels can be different in some
instances, while in some others they are virtually identical
\cite{armitage10,stadlober95,kendziora01,blumberg02,qazilbash05}. In
this case, the investigating similarities and differences of ERR
between the $p$-doped and $n$-doped cuprate superconductors would be
crucial to understanding physics of superconductivity in doped
cuprates.

Theoretically, ERR has been thoroughly studied based on the
phenomenological Bardeen-Cooper-Schrieffer (BCS) formalism with a
monotonic d-wave gap and the microscopic model and understood in the
$p$-doped cuprate superconductors
\cite{devereaux07,zeyher02,devereaux94,branch95,manske97,dahm98,misochko00,chubukov06},
however, no similar studies have been made so far for the $n$-doped
counterparts. In this case, an important issue is what aspects of
the low energy excitations determined by ERR in both $p$-doped and
$n$-doped cuprate superconductors are universal. In our recent work
\cite{geng10} based on the kinetic energy driven SC mechanism
\cite{feng03}, the doping and temperature dependence of ERR in the
$p$-doped cuprate superconductors has been discussed, and some main
features of ERR in the $p$-doped cuprate superconductors are
qualitatively reproduced
\cite{chen98,guyard08,venturini02,guyard08b,tacon06}. In this paper,
we study the doping and temperature dependence of ERR in the
$n$-doped counterparts along with this line. We show explicitly that
the pair-breaking peak energy in the $B_{2g}$ symmetry continuously
follows the SC transition temperature $T_{c}$ throughout the SC dome
as $\omega^{B_{2g}}_{{\rm peak}} \propto T_{c}$, and therefore is a
common feature for both $n$-doped and $p$-doped cuprate
superconductors. However, there are pronounced deviations from a
cubic response in the $B_{1g}$ channel and a linear response in the
$B_{2g}$ channel in the low energies. Our results also show that
these pronounced deviations are intriguingly related to a
nonmonotonic d-wave gap in the $n$-doped cuprate superconductors.

It has been shown that the essential physics of the $n$-doped
cuprate superconductors is contained in the $t$-$t'$-$J$ model on a
square lattice \cite{damascelli03,anderson87},
\begin{eqnarray}\label{t-jmodel}
H&=&t\sum_{i\hat{\eta}\sigma}PC^{\dag}_{i\sigma}
C_{i+\hat{\eta}\sigma}P^{\dag}-t'\sum_{i\hat{\tau}\sigma}
PC^{\dag}_{i\sigma}C_{i+\hat{\tau}\sigma}P^{\dag}\nonumber\\
&-&\mu\sum_{i\sigma}
PC^{\dag}_{i\sigma}C_{i\sigma} P^{\dag}+J\sum_{i\hat{\eta}}{\bf
S}_i\cdot{\bf S}_{i+\hat{\eta}},
\end{eqnarray}
where $t<0$, $t'<0$, $\hat{\eta}=\pm\hat{x},\pm\hat{y}$,
$\hat{\tau}=\pm\hat{x} \pm\hat{y}$, $C^{\dagger}_{i\sigma}$
($C_{i\sigma}$) is the electron creation (annihilation) operator,
${\bf S}_{i}=(S_{i}^{x},S_{i}^{y},S_{i}^{z})$ are spin operators,
and $\mu$ is the chemical potential. The nontrivial part of the
$t$-$t'$-$J$ model (\ref{t-jmodel}) resides in the projection
operator $P$ which restricts the Hilbert space to exclude the zero
occupancy, i.e., $\sum_{\sigma}C^{\dagger}_{i\sigma}C_{i\sigma}\geq
1$. In this case, an important question is the relation between the
$p$-doped and $n$-doped cuprate superconductors. The $t$-$J$ model
with nearest neighbor hopping $t$ has a particle-hole symmetry
because the sign of $t$ can be absorbed by changing the sign of the
orbital on one sublattice. However, the particle-hole asymmetry can
be described by including the next-nearest neighbor hopping $t'$
\cite{hybertson90,pavarini01}. Although there is a similar strength
of the magnetic interaction $J$ for both $p$-doped and $n$-doped
cuprate superconductors, the interplay of $t'$ with $t$ and $J$
causes a further weakening of the AF spin correlation for the
$p$-doped cuprate superconductors, and enhancing the AF spin
correlation in the $n$-doped counterparts
\cite{hybertson90,pavarini01,gooding94}, therefore the AF spin
correlation in the $n$-doped case is stronger than that in the
$p$-doped side.

For incorporating the single occupancy constraint in the $p$-doped
cuprate superconductors, we have developed a charge-spin separation
(CSS) fermion-spin theory \cite{feng04}. For description of the
$n$-doped counterparts within this CSS fermion-spin theory, the
$t$-$t'$-$J$ model (\ref{t-jmodel}) can be rewritten in terms of a
particle-hole transformation $C_{i\sigma} \rightarrow
f^{\dagger}_{i-\sigma}$ as,
\begin{eqnarray}\label{t-jmodel1}
H&=&-t\sum_{i\hat{\eta}\sigma}f^{\dag}_{i\sigma}
f_{i+\hat{\eta}\sigma}+t'\sum_{i\hat{\tau}\sigma}
f^{\dag}_{i\sigma}f_{i+\hat{\tau}\sigma}\nonumber\\
&+&\mu\sum_{i\sigma}
f^{\dag}_{i\sigma}f_{i\sigma} +J\sum_{i\hat{\eta}}{\bf S}_i\cdot{\bf
S}_{i+\hat{\eta}},
\end{eqnarray}
then the local constraint $\sum_{\sigma}C^{\dagger}_{i\sigma}
C_{i\sigma}\geq 1$ is transferred as $\sum_{\sigma}
f^{\dagger}_{i\sigma}f_{i\sigma}\leq 1$, where
$f^{\dagger}_{i\sigma}$ ($f_{i\sigma}$) is the hole creation
(annihilation) operator. Now we follow the CSS fermion-spin theory,
and decouple the hole operators as $f_{i\uparrow}=
a^{\dagger}_{i\uparrow}S^{-}_{i}$ and $f_{i\downarrow}=
a^{\dagger}_{i\downarrow}S^{+}_{i}$, respectively, where the spinful
fermion operator $a_{i\sigma}=e^{-i\Phi_{i\sigma}}a_{i}$ represents
the charge degree of freedom together with some effects of the spin
configuration rearrangements due to the presence of the doped
electron itself (charge carrier), while the spin operator $S_{i}$
represents the spin degree of freedom, then the single occupancy
local constraint $\sum_{\sigma} f^{\dagger}_{i\sigma}f_{i\sigma}\leq
1$ is satisfied in analytical calculations. In this CSS fermion-spin
representation, the $t$-$t'$-$J$ model (\ref{t-jmodel1}) can be
expressed as,
\begin{eqnarray}\label{t-jmodel2}
H&=&t\sum_{i\hat{\eta}}(a^{\dagger}_{i+\hat{\eta}\uparrow}a_{i\uparrow}
S^{+}_{i}S^{-}_{i+\hat{\eta}}+a^{\dagger}_{i+\hat{\eta}\downarrow}
a_{i\downarrow}S^{-}_{i}S^{+}_{i+\hat{\eta}})\nonumber\\
&-&t'\sum_{i\hat{\tau}}
(a^{\dagger}_{i+\hat{\tau}\uparrow}a_{i\uparrow}S^{+}_{i}
S^{-}_{i+\hat{\tau}}+a^{\dagger}_{i+\hat{\tau}\downarrow}
a_{i\downarrow}S^{-}_{i}S^{+}_{i+\hat{\tau}}) \nonumber \\
&-&\mu\sum_{i\sigma}a^{\dagger}_{i\sigma}a_{i\sigma}+J_{{\rm eff}}
\sum_{i\hat{\eta}}{\bf S}_{i}\cdot {\bf S}_{i+\hat{\eta}},
\end{eqnarray}
where $J_{\rm {eff}}=(1-\delta)^2J$, and $\delta=\langle
a^{\dag}_{i\sigma}a_{i\sigma}\rangle=\langle a^{\dag}_{i}a_{i}
\rangle$ is the electron doping concentration. As in the discussions
of the $p$-doped case \cite{feng03}, the SC order parameter for the
electron Cooper pair in the $n$-doped side also can be defined as,
\begin{eqnarray}
\Delta &=&\langle C^{\dag}_{i\uparrow}C^{\dag}_{j\downarrow}-
C^{\dag}_{i\downarrow}C^{\dag}_{j\uparrow}\rangle=\langle
a_{i\uparrow}a_{j\downarrow}S^{\dag}_{i}S^{-}_{j}-
a_{i\downarrow}a_{j\uparrow}S^{-}_{i}S^{+}_{j}\rangle\nonumber\\
&=&-\langle
S^{+}_{i}S^{-}_{j} \rangle\Delta_{a},
\end{eqnarray}
with the charge carrier pairing order parameter $\Delta_{a}=\langle
a_{j\downarrow} a_{i\uparrow}-a_{j\uparrow}a_{i\downarrow}\rangle$.
It has been shown experimentally that the hot spots are located
close to $[\pm\pi,0]$ and $[0,\pm\pi]$ (in units of inverse lattice
constant) points of the Brillouin zone (BZ) in the $p$-doped case,
resulting in a monotonic d-wave gap \cite{ding96}. In contrast, the
hot spots are located much closer to the zone diagonal in the
$n$-doped side, leading to a nonmonotonic d-wave gap
\cite{matsui05,blumberg02},
\begin{eqnarray}\label{gap}
\Delta({\bf k})=\Delta[\gamma^{(d)}_{{\bf k}}-B\gamma^{(2d)}_{{\bf k
}}],
\end{eqnarray}
where $\gamma^{(d)}_{{\bf k}}=[{\rm cos}k_{x}-{\rm cos}k_{y}]/2$ and
$\gamma^{(2d)}_{{\bf k}}=[{\rm cos}(2k_{x})-{\rm cos}(2k_{y})] /2$,
then the maximum gap is observed not at the BZ boundary as expected
from the monotonic d-wave gap, but at the hot spot between [$\pi$,0]
and [$\pi/2$,$\pi/2$], where the AF spin fluctuation most strongly
couples to electrons, supporting a spin-mediated pairing mechanism
\cite{matsui05}.

For a microscopic description of the SC state in cuprate
superconductors, we \cite{feng03} have developed a kinetic energy
driven SC mechanism, where the interaction between charge carriers
and spins from the kinetic energy term in the $t$-$t'$-$J$ model
(\ref{t-jmodel2}) induces a charge carrier pairing state with the
d-wave symmetry by exchanging spin excitations, while the electron
Cooper pairs originating from this charge carrier pairing state are
due to the charge-spin recombination, and their condensation reveals
the SC ground-state. Moreover, this SC state is a conventional
BCS-like with the d-wave symmetry \cite{guo07}, so that some main
features of the low energy electronic structure of both $p$-doped
and $n$-doped cuprate superconductors have been quantitatively
reproduced \cite{guo07,cheng07}. In particular, the doping and
energy evolution of the magnetic excitations in the $n$-doped
cuprate superconductors has been studied \cite{cheng08} in terms of
a nonmonotonic d-wave gap (\ref{gap}), and the results show that
there is a broad commensurate scattering peak at low energy, then
the resonance energy is located among this low energy commensurate
scattering range. This low energy commensurate scattering disperses
outward into a continuous ring-like incommensurate scattering at
high energy. Following our previous discussions
\cite{cheng07,cheng08}, the full charge carrier Green function of
the $n$-doped cuprate superconductors can be obtained in the Nambu
representation as,
\begin{eqnarray}\label{holegreenfunction}
g({\bf{k}},i\omega_n)=Z_{\rm{aF}}\,\frac{i\omega_n\tau_0 +
\bar{\xi}_{{\rm a}\bf{k}}\tau_3 - \bar{\Delta}_{\rm{aZ}}({\bf{k}})
\tau_1}{(i\omega_n)^2 - E_{{\rm{a}}{\bf{k}}}^2},
\label{holegreenfunction}
\end{eqnarray}
where $\tau_{0}$ is the unit matrix, $\tau_{1}$ and $\tau_{3}$ are
Pauli matrices, the renormalized charge carrier excitation spectrum
$\bar{\xi}_{{\rm a}\bf k}=Z_{\rm aF}\xi_{{\rm a}\bf k}$, with the
mean-field charge carrier excitation spectrum $\xi_{{\rm a}\bf k}=
Zt\chi_{1} \gamma^{(s)}_{{\bf k}}-Zt'\chi_{2}\gamma^{(2s)}_{{\bf
k}}-\mu$, the spin correlation functions $\chi_{1}=\langle S_{i}^{+}
S_{i+\hat{\eta}}^{-} \rangle$ and $\chi_{2}= \langle S_{i}^{+}
S_{i+\hat{\tau}}^{-}\rangle$, $\gamma^{(s)}_{\bf k}=(1/Z)
\sum_{\hat{\eta}}e^{i{\bf k}\cdot\hat{\eta}}$, $\gamma^{(2s)}_{\bf
k}= (1/Z)\sum_{\hat{\tau}}e^{i{\bf k}\cdot\hat{\tau}}$, $Z$ is the
number of the nearest neighbor or next-nearest neighbor sites, the
renormalized charge carrier d-wave pair gap $\bar{\Delta}_{\rm aZ}
({\bf k})=Z_{\rm aF}\bar{\Delta}_{\rm a}({\bf k})$, with the
effective charge carrier d-wave pair gap $\bar{\Delta}_{\rm a}({\bf
k})=\bar{\Delta}_{\rm a}[\gamma^{(d)}_{{\bf k}}-B\gamma^{(2d)}_{{\bf
k }}]$, and the charge carrier quasiparticle spectrum $E_{{\rm a}
{\bf k}}=\sqrt{\bar{\xi}^{2}_{{\rm a}{\bf k}}+|\bar{\Delta}_{\rm
aZ}({\bf k} ) |^{2}}$, while the effective charge carrier pair gap
$\bar{\Delta}_{\rm a}({\bf k})$ and the quasiparticle coherent
weight $Z_{\rm aF}$ satisfy the following equations \cite{cheng08}
$\bar{\Delta}_{\rm a}({\bf k})=\Sigma^{(a)}_{2}({\bf k},\omega=0)$
and $Z^{-1}_{\rm aF}=1-\Sigma^{(a)}_{1{\rm o}}({\bf k}, \omega=0)
\mid_{{\bf k}=[\pi,0]}$, where $\Sigma^{(a)}_{1}({\bf k},\omega)$
and $\Sigma^{(a)}_{2}({\bf k},\omega)$ are the charge carrier
self-energies obtained from the spin bubble, and have been given in
Ref. \onlinecite{cheng07} except the effective charge carrier
monotonic d-wave gap has been replaced by the present nonmonotonic
one, while $\Sigma^{(a)}_{1{\rm o}}({\bf k},\omega)$ is the
antisymmetric part of $\Sigma^{(a)}_{1}({\bf k},\omega)$. These
equations have been solved simultaneously with other self-consistent
equations \cite{cheng07,cheng08}, then all order parameters,
decoupling parameter $\alpha$, and chemical potential $\mu$ have
been determined by the self-consistent calculation.

In the CSS fermion-spin representation \cite{feng04}, the electron
Green's function,
\begin{eqnarray}
G({\bf k}, i\omega_n)=\left(
\begin{array}{cccc}
G_{11}({\bf k},i\omega_n), & G_{12}({\bf k},i\omega_n) \\
G_{21}({\bf k},i\omega_n), & G_{22}({\bf k},i\omega_n)
\end{array} \right) \,, \label{electrongreenfunction}
\end{eqnarray}
is a convolution of the spin Green's function and charge carrier
Green's function (\ref{holegreenfunction}), and its diagonal and
off-diagonal components $G_{11} (i-j,t-t') =\langle\langle
C_{i\sigma}(t); C^{\dagger}_{j\sigma} (t')\rangle \rangle$ and
$G_{21}(i-j,t-t')=\langle \langle C^{\dagger}_{i\uparrow}(t);
C^{\dagger}_{j\downarrow}(t')\rangle \rangle$ have been given in
Ref. \cite{cheng07}.

For discussions of ERR in the $n$-doped cuprate superconductors, we
need to calculate the ERR function $\tilde{S}({\bf q},\omega)$,
which can be obtained in terms of the imaginary part of the Raman
density-density correlation function $\tilde{\chi}({\bf q},\omega)$
as \cite{devereaux07},
\begin{eqnarray}
\tilde{S}({\bf q},\omega)=-{1\over\pi}[1+n_{B}(\omega)]{\rm Im}
\tilde{\chi}({\bf q},\omega), \label{ramanscattering}
\end{eqnarray}
with $n_{B}(\omega)$ is the boson distribution function, while the
Raman density-density correlation function $\tilde{\chi}({\bf q},
\omega)$ is defined as $\tilde{\chi}({\bf q},\tau-\tau')=-\langle
T\rho_{\gamma}({\bf q}, \tau) \rho_{\gamma}(-{\bf q},\tau')\rangle$,
with the Raman density operator in the Nambu representation
$\rho_{\gamma}({\bf q})=\sum_{{\bf k}}\gamma_{\bf k}
C^{\dagger}_{{\bf k}+{{\bf q}\over 2}}\tau_{3} C_{{\bf k}-{{\bf q}
\over 2}}$, where the bare Raman vertex $\gamma_{\bf k}$ can be
classified by the representations $B_{1g}$, $B_{2g}$, and $A_{1g}$
of the point group $D_{4h}$ of the square lattice as
\cite{devereaux07},
\begin{eqnarray}
\gamma_{\bf k}=\left\{
\begin{array}{ll}
b_{\omega_{i},\omega_{s}}\left[\cos(k_{x}a)-\cos(k_{y}a)
\right]/4,& B_{1g},\\
b'_{\omega_{i},\omega_{s}}\sin(k_{x}a)\sin(k_{y}a),& B_{2g},\\
a_{\omega_{i},\omega_{s}}\left[\cos(k_{x}a)+\cos(k_{y}a)\right]/4,&
A_{1g},
\end{array}
\right. \label{vertex}
\end{eqnarray}
respectively, where as a qualitative discussion, the magnitude of
the energy dependence of the prefactors $b$, $b'$ and $a$ can be
rescaled to units. The Raman vertex $\gamma_{\bf k}$ (\ref{vertex})
also shows that for the $B_{2g}$ channel, the Raman vertex is most
sensitive to the nodal ($[\pm\pi/2,\pm\pi/2]$) region of the BZ and
vanishes along $[0,0]\rightarrow [\pi,0]$ and equivalent lines of
the BZ. However, for the $B_{1g}$ channel, the nodal
$[0,0]\rightarrow [\pi,\pi]$ diagonals do not contribute to the
intensity that mainly integrates from regions near intersections of
the Fermi surface and the BZ boundary. In this case, the Raman
density-density correlation function $\tilde{\chi}({\bf q},\omega)$
can be expressed in terms of the electron Green's function
(\ref{electrongreenfunction}) as,
\begin{eqnarray}\label{ramancorrelationfunction}
\tilde{\chi}_{\gamma_{1}\gamma_{2}}({\bf q},iq_{n})&=&{1\over N}
\sum_{{\bf k}}\gamma_{1{\bf k}}\gamma_{2{\bf k}}{1\over\beta}
\sum_{i\omega_{n}}{\rm Tr}[G({\bf k}-{\bf q}/2,i\omega_{n})
\nonumber\\
&\times&\tau_{3}G({\bf k}+{\bf q}/2,i\omega_{n}-iq_{n})\tau_{3}],
\end{eqnarray}
then the explicit form of the obtained
$\tilde{\chi}_{\gamma_{1}\gamma_{2}}({\bf q},\omega)$ in the present
$n$-doped cuprate superconductors is the same as that given in Ref.
\onlinecite{geng10} for the $p$-doped counterparts except the charge
carrier quasiparticle coherence factors $U^{2}_{{\rm 1h}{\bf k}}=
(1+\bar{\xi_{{\bf k}}} /E_{{\rm h}{\bf k}})/2$ and $U^{2}_{{\rm 2h}
{\bf k}}=(1- \bar{\xi_{{\bf k}}}/E_{{\rm h}{\bf k}})/2$, the
renormalized charge carrier d-wave pair gap $\bar{\Delta}_{\rm hZ}
({\bf k})$, and the charge carrier quasiparticle spectrum $E_{{\rm
h} {\bf k}}$ have been replaced as $U^{2}_{{\rm 1a}{\bf k}}=(1-
\bar{\xi_{{\bf k}}}/ E_{{\rm a}{\bf k}}) /2$ and $U^{2}_{{\rm 2a}
{\bf k}}=(1+ \bar{\xi_{{\bf k}}}/E_{{\rm a} {\bf k}})/2$,
$\bar{\Delta}_{\rm aZ} ({\bf k})$, and $E_{{\rm a} {\bf k}}$,
respectively. This result shows that the Raman density-density
correlation function (\ref{ramancorrelationfunction}) [then the ERR
function (\ref{ramanscattering})] is closely related to the pair
coherence factors, which give rise to the occurrence of the
pair-breaking features in the spectra \cite{devereaux07,geng10}. The
above Raman density-density correlation function
(\ref{ramancorrelationfunction}) also shows that the variation with
momentum of the Raman vertex $\gamma_{{\bf k}}$ is coupled to the
momentum dependence of the charge carrier nonmonotonic d-wave gap
$\bar{\Delta}_{{\rm aZ}}({\bf k})$ in the $n$-doped case, leading to
a strong polarization dependence of the spectra.

\begin{figure}[h!]
\includegraphics[scale=0.45]{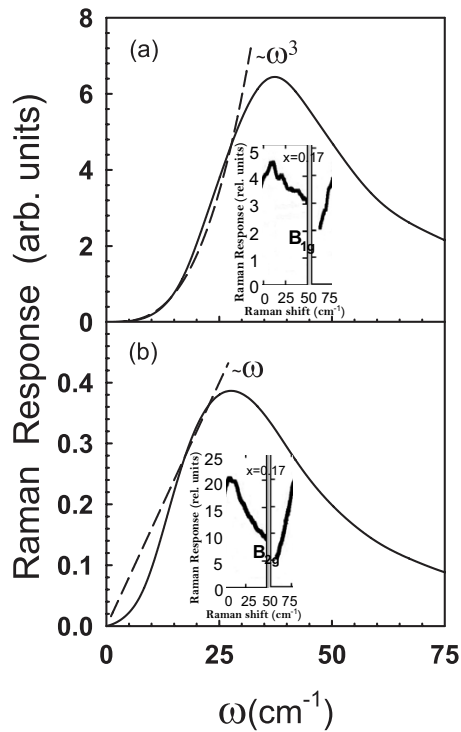}
\caption{(a) $B_{1g}$ and (b) $B_{2g}$ spectra as a function of
energy at $p=0.17$ for $T=3$K. The dashed lines are a cubic and a
linear fit for the low-energy $B_{1g}$ and $B_{2g}$ spectra,
respectively. Inset: the corresponding experimental results of
Pr$_{2-x}$Ce$_{x}$CuO$_{4-\delta}$ taken from Ref.
\onlinecite{qazilbash05}.} \label{fig1}
\end{figure}

For the $n$-doped cuprate superconductors, although the values of
$J$, $t$, and $t'$ in the $t$-$t'$-$J$ model (\ref{t-jmodel}) are
believed to vary somewhat from compound to compound, the numerical
calculations \cite{hybertson90,pavarini01} have extracted the range
of these parameters as $J\approx 0.1\sim 0.13$eV, $t/J=-2.5\sim -3$,
and $t'/t$ is of order $0.2\sim 0.3$. However, as a qualitative
discussion in this paper, the commonly used parameters are chosen as
$t/J=-2.5$, $t'/t=0.3$, and $J=0.13$eV$\approx 1500$K$\approx
1049$cm$^{-1}$. We are now ready to discuss the doping and
temperature dependence of ERR in the $n$-doped cuprate
superconductors. We have performed a calculation for the ERR
function (\ref{ramanscattering}) in both $B_{1g}$ and $B_{2g}$
orientations, and the results of (a) the $B_{1g}$ and (b) $B_{2g}$
spectra at the overdoping $p=0.17$ in temperature $T=3$K are plotted
in Fig. \ref{fig1} in comparison with the corresponding experimental
results \cite{qazilbash05} of the $n$-doped cuprate superconductor
Pr$_{2-x}$Ce$_{x}$CuO$_{4-\delta}$ (PCCO) are also presented in Fig.
\ref{fig1} (inset). As in the hole-doped case \cite{geng10}, both
$B_{1g}$ and $B_{2g}$ spectra in the overdoped regime are
characterized by the presence of the pair-breaking $(2\Delta)$ peaks
and low-energy tails. These low-energy tails of the Raman continuum
change to reflect the opening of the gap, i.e., the strength of the
low-energy continuum is reduced and the spectrum acquires the
pair-breaking peak as a result of excitations across the anisotropic
gap $2\Delta({\bf k})$, while the pair-breaking peaks correspond to
the excitations out of the SC condensate. However, for different
scattering geometries spectra differ in their intensity as well as
in the position of the $2\Delta$ peaks. Although the effective
charge carrier pair gap $\bar{\Delta}_{{\rm aZ}}({\bf k})$ [then the
SC gap (\ref{gap})] obviously deviates from the monotonic d-wave gap
\cite{matsui05}, it is basically consistent with the d-wave
symmetry. This leads to that the pair-breaking peak intensities in
both $B_{1g}$ and $B_{2g}$ below the SC coherence peaks vanish
smoothly without a threshold to zero energy \cite{geng10}.
Furthermore, we have also fitted the low-energy tails to an
$\omega^{3}$ power law in the $B_{1g}$ response and linearly with
$\omega$ in the $B_{2g}$ response, however, in sharp contrast to the
$p$-doped case \cite{geng10}, there are pronounced deviations from a
cubic response in the $B_{1g}$ channel and a linear response in the
$B_{2g}$ channel. For the $p$-doped cuprate superconductors, the
interpretation of the cubic response in the $B_{1g}$ channel and
linear response in the $B_{2g}$ channel is consistent with the
monotonic d-wave gap \cite{devereaux07,geng10}. In comparison with the
$p$-doped case, our present results also show that these pronounced
deviations in the low energies in the $n$-doped counterparts are
mainly caused by the nonmonotonic d-wave gap, which is also
consistent with the interpretation in terms of the phenomenological
BCS formalism with the nonmonotonic d-wave gap given in Ref.
\onlinecite{blumberg02}. However, there is a substantial difference
between theory and experiment, namely, in the extremely low energies,
the calculated ERR spectrum in the $B_{1g}$ channel is roughly
consistent with an $\omega^{3}$ power law, while the calculated ERR
spectrum in the $B_{2g}$ channel is in disagreement with a linear
$\omega$ response. This result in the extremely low energies seems not
to be consistent with the corresponding experimental data
\cite{qazilbash05}, where in the extremely low energies, the ERR
spectrum in the $B_{1g}$ channel does not show an $\omega^{3}$ behavior,
while the ERR spectrum in the $B_{2g}$ channel may be a linear $\omega$
dependence. The simple $t$-$J$ model can not be regarded as a complete
model for the quantitative comparison with cuprate superconductors,
however, as a qualitative discussion in this paper, the overall shape
seen in the theoretical result is qualitatively consistent with that
observed in the experiment \cite{qazilbash05}.

\begin{figure}[h!]
\includegraphics[scale=0.45]{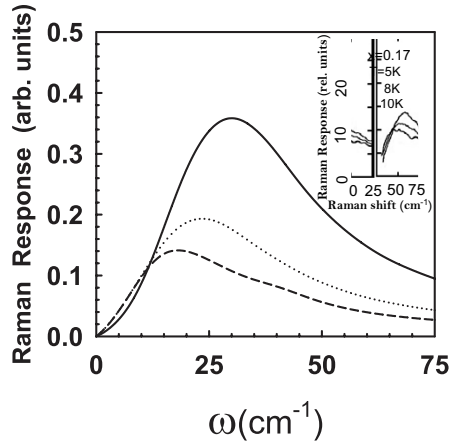}
\caption{$B_{2g}$ spectrum as a function of energy at $p=0.17$ for
$T=5.2$K (solid line), $T=8.2$K (dotted line), and $T=10.5$K (dashed
line). Inset: the corresponding experimental results of
Pr$_{2-x}$Ce$_{x}$CuO$_{4-\delta}$ taken from Ref.
\onlinecite{qazilbash05}.} \label{fig2}
\end{figure}

\begin{figure}[h!]
\includegraphics[scale=0.45]{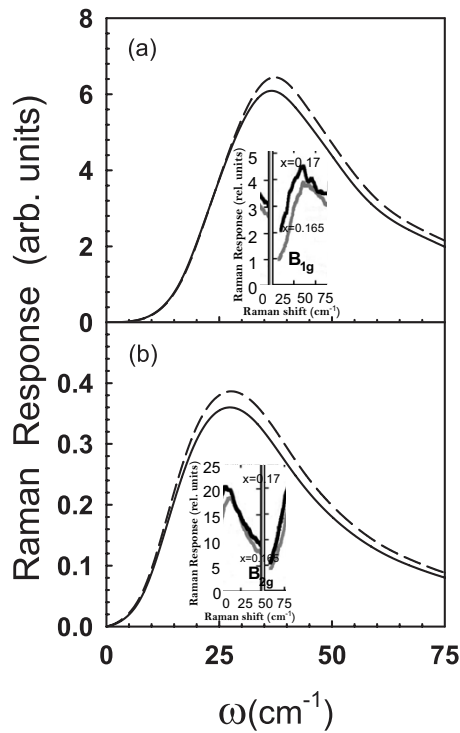}
\caption{(a) $B_{1g}$ and (b) $B_{2g}$ spectra as a function of
energy at $p=0.165$ (solid line) and $p=0.17$ (dashed line) for
$T=3$K. Inset: the corresponding experimental results of
Pr$_{2-x}$Ce$_{x}$CuO$_{4-\delta}$ taken from Ref.
\onlinecite{qazilbash05}.} \label{fig3}
\end{figure}

ERR in Fig. \ref{fig1} is also temperature dependent. To analyze the
evolution of ERR in Fig. \ref{fig1} with temperature, we have
performed a calculation for the ERR function (\ref{ramanscattering})
in the $B_{2g}$ channel with different temperatures, and the results
of the $B_{2g}$ spectrum as a function of energy with $T=5.2$K
(solid line), $T=8.2$K (dotted line), and $T=10.5$K (dashed line)
for $p=0.17$ are plotted in Fig. \ref{fig2} in comparison with the
corresponding experimental results \cite{qazilbash05} of PCCO
(inset). Within the kinetic energy driven SC mechanism, the calculated
SC transition temperature $T_{c}=33$K at $p=0.17$. This anticipated SC
transition temperature is not too far from the SC transition temperature
$T_{c}=13$K observed experimentally on PCCO
at $p=0.17$ \cite{qazilbash05}. Obviously, the experimental results
of the ERR spectrum \cite{qazilbash05} are qualitatively reproduced.
In particular, the pair-breaking peak loses intensity and moves to
lower energies by increasing temperatures. Furthermore, as in the
$p$-doped case \cite{geng10}, this pair-breaking peak intensity in
the $n$-doped counterparts also follows a pair gap type temperature
dependence, and disappears at the SC transition temperature $T_{c}$.

\begin{figure}[h!]
\includegraphics[scale=0.45]{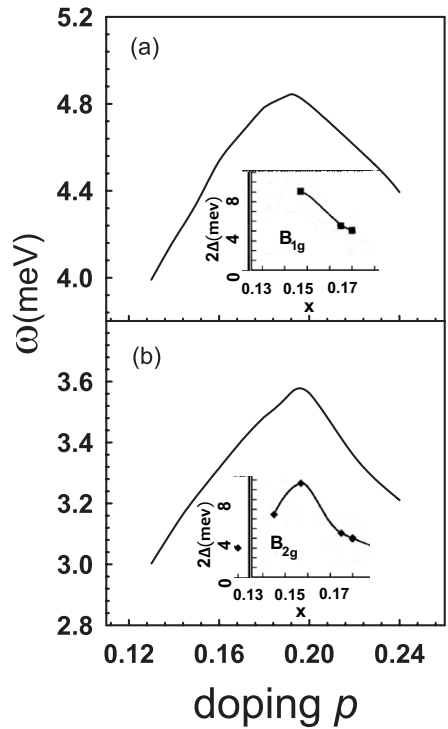}
\caption{(a) $B_{1g}$ and (b) $B_{2g}$ peaks as a function of doping
for $T=3$K. Inset: the corresponding experimental results of
Pr$_{2-x}$Ce$_{x}$CuO$_{4-\delta}$ taken from Ref.
\onlinecite{qazilbash05}.} \label{fig4}
\end{figure}

Now we turn to discuss the doping dependence of ERR. We have made a
series of calculations for the ERR function (\ref{ramanscattering})
with different doping concentrations, and the results of the (a)
$B_{1g}$ and (b) $B_{2g}$ spectra as a function of energy with
$p=0.165$ (solid line) and $p=0.17$ (dashed line) at $T=3$K are
plotted in Fig. \ref{fig3} in comparison with the corresponding
experimental results \cite{qazilbash05} of PCCO (inset). Our results
show that with increasing the doping concentration in the overdoped
regime, the weights of the peaks in both $B_{1g}$ and $B_{2g}$
symmetries increase, while the peak energies in both $B_{1g}$ and
$B_{2g}$ channels decrease. However, since there is the weakness of
the calculated ERR spectrum in the extremely low energies as shown in
Fig. \ref{fig1}, the doping dependence of the ERR spectrum in the low
energies is better than that in the extremely low energies. Moreover,
in contrast to the case in the overdoped regime, the peak energies in
both $B_{1g}$ and $B_{2g}$ channels move to the higher energies when
increasing doping concentration in the underdoped regime. To show this
point clearly, we have calculated the ERR function (\ref{ramanscattering})
throughout the SC dome, and then employed the shift of the
leading-edge mid-point as a measurement of the magnitude of the gap
at each doping concentration just as it has been done in the
experiments \cite{qazilbash05}. The results for the extracted (a)
$B_{1g}$ and (b) $B_{2g}$ peak energies as a function of doping with
$T=3$K are plotted in Fig. \ref{fig4}. For comparison, the
corresponding experimental results \cite{qazilbash05} of PCCO are
also presented in Fig. \ref{fig4} (inset). It is shown that both
peak energies have pronounced maximum around the optimal doping, and
then decreases in both underdoped and overdoped regimes. The maximum
values of the peak energy for the $B_{2g}$ channel throughout the SC
dome are very similar to the single particle spectroscopy gap values
\cite{damascelli03,cheng07}, and therefore continuously follows the
SC transition temperature as $\omega^{B_{2g}}_{{\rm peak}}\propto
T_{c}$. In this sense, it can be closely associated with twice the
gap magnitude, in good agreement with the experimental results
\cite{qazilbash05}. Furthermore, incorporating with our previous
results of ERR for the $p$-doped case \cite{geng10}, our present
results also show that the domelike shape of the doping dependent
peak energy in the $B_{2g}$ symmetry is a common feature for both
electron-doped and hole-doped cuprate superconductors. In the
$B_{1g}$ orientation, our present results of the maximum values of
the peak energy in the optimally and overdoped regimes are in
qualitative agreement with the the corresponding experimental
results \cite{qazilbash05} of PCCO. This seems to show that peak
energy in $B_{1g}$ channel in the optimally and overdoped regimes is
qualitatively related to the gap. However, in the underdoped regime,
no extracted maximum values of the Raman scattering peak energy from
experimental measurements on the $n$-doped counterparts are
available now. In this case, we can not make a direct comparison for
the present theoretical results with the corresponding experimental
data in the underdoped regime, and therefore present theoretical
results in the underdoped regime need to be verified by further
Raman scattering experiments. On the other hand, since the integrated
reduced coherence peak intensity in the SC state is proportional to
the superfluid density $\rho_{s}$ \cite{qazilbash05}, therefore the
domelike shape of the doping dependent ERR spectrum in the $n$-doped
cuprate superconductors implies that the superfluid density $\rho_{s}$
increases with increasing doping in the lower doped regime, and
reaches a maximum around the critical doping, then decreases in the
higher doped regime. This has been confirmed by our study of the
doping dependence of the superfluid density in the $n$-doped cuprate
superconductors \cite{huang12}.

The electronic states in solids are characterized by their energy
dispersions as well as the characteristic lifetime of an electron
placed into such a state. This state is represented by the
single-particle propagator, while the spectral function is directly
related to the analytically continued single-particle propagator
\cite{devereaux07}. In particular, this spectral function is
measurable via angle-resolved photoemission spectroscopy (ARPES)
techniques and can provide an important information about
quasiparticles \cite{damascelli03}. Within the framework of the
kinetic energy driven SC mechanism, the essential physics of ERR in
the $n$-doped cuprate superconductors is the same as that in the
$p$-doped counterparts \cite{geng10}. The kinetic energy driven
SC-state in the $n$-doped cuprate superconductors is the
conventional BCS like with the d-wave symmetry \cite{cheng07}, where
it has been shown that this kinetic energy driven d-wave BCS
formalism can reproduce quantitatively some main low energy features
of the ARPES experimental data of the $n$-doped cuprate
superconductors. However, many of the unusual physical properties of
the two-particle electron dynamics across the phase diagram are
attributed to particular characteristics of low energy excitations
determined by ERR. Although the differences between single-particle
and two-particle properties are inescapable in cuprate
superconductors, the single-particle and two-particle correlation
functions can be related each other \cite{devereaux07}. This is why
the kinetic energy driven d-wave BCS formalism is still valid in
discussions of the doping and temperature dependence of ERR in the
$n$-doped cuprate superconductors.

In the $p$-doped cuprate superconductors, the Raman scattering
experimental data show that the peak energy in $B_{1g}$ channel
increases essentially linearly as the doping concentration is
reduced in the underdoped regime \cite{guyard08b,tacon06}, which
reflects that the peak energy in the $B_{1g}$ symmetry is
progressively disconnected from superconductivity as one goes from
the overdoped regime to the underdoped regime. In Ref.
\onlinecite{tacon06}, it has been argued that for a correct
description of the $B_{1g}$ spectrum in the $p$-doped cuprate
superconductors in the underdoped regime, two essential ingredients
should be taken into account within the d-wave BCS formalism: (1)
the quasiparticle spectral weight $Z_{{\rm F}}({\bf k})$ as well as
the vertex correction describing the interaction of the
quasiparticles with external perturbations; and (2) a general gap
also should be taken into account by including the higher order of
the harmonic component in the monotonic d-wave gap. In our recent
work based on the monotonic d-wave gap for the $p$-doped cuprate
superconductors \cite{geng10}, the first condition is partially
satisfied, since the doping dependence of the quasiparticle spectral
weight $Z_{{\rm F}}$ has been included in the discussions. However,
the vertex correction for the interaction of the quasiparticles due
to the presence of the spin fluctuation is not included, which also
is doping and momentum dependent. In this case, our theoretical
results \cite{geng10} of ERR for the $p$-doped cuprate
superconductors in the $B_{1g}$ channel in the underdoped regime is
in disagreement with the corresponding experimental data
\cite{guyard08b,tacon06}. For the second condition, our present
study shows indirectly that if the vertex correction for the
interaction of the quasiparticles with external perturbations is not
considered, even a more general nonmonotonic d-wave gap (\ref{gap})
is used for discussions of ERR in the $p$-doped cuprate
superconductors, the correct behavior of ERR in the $B_{1g}$ channel
in the underdoped regime still can not be reproduced. Thus an
important issue is how the vertex correction for the interaction of
the quasiparticles due to the presence of the spin fluctuation is
taken into account within the kinetic energy driven d-wave BCS
formalism for a correct description of the $B_{1g}$ spectrum of the
$p$-doped cuprate superconductors in the underdoped regime. On the
other hand, the effect of a weak magnetic field on ERR is not
considered in the present work. However, the depairing due to the
Pauli spin polarization is very important in the presence of a weak
magnetic field, since both $p$-doped and $n$-doped cuprate
superconductors are doped Mott insulators with the strong short-range
AF correlation dominating the entire SC phase \cite{damascelli03}. In
particular, in the kinetic energy driven SC mechanism \cite{feng03},
where the charge carrier-spin interaction from the kinetic energy term
induces a d-wave pairing state by exchanging spin excitations.
Therefore under the kinetic energy driven SC mechanism, an external
magnetic field aligns the spins of the unpaired electrons, then the
d-wave electron Cooper pairs in both $p$-doped and $n$-doped cuprate
superconductors can not take advantage of the lower energy offered by
a spin-polarized state. This leads to that a weak magnetic field may
suppress the pair-breaking peak magnitude in the ERR spectrum at a
rapid rate \cite{qazilbash05,huang11}. These and the related issues
are under investigation now.

In summary, we have studied the doping and temperature dependence of
ERR in the $n$-doped cuprate superconductors based on the kinetic
energy driven SC mechanism. Our results show that the pair-breaking
peak energy in the $B_{2g}$ symmetry continuously follows the SC
transition temperature $T_{c}$ throughout the SC dome as
$\omega^{B_{2g}}_{{\rm peak}} \propto T_{c}$, and therefore is a
common feature for both $n$-doped and $p$-doped cuprate
superconductors. However, in sharp contrast to the $p$-doped
counterparts, there are pronounced deviations from a cubic response
in the $B_{1g}$ channel and a linear response in the $B_{2g}$
channel in the low energies. Our results also show that these
pronounced deviations are mainly caused by a nonmonotonic d-wave gap
in the $n$-doped cuprate superconductors.

Within the framework of the kinetic energy driven SC mechanism, the
electromagnetic response in both $p$-doped and $n$-doped cuprate
superconductors in the SC state has been studied \cite{huang11,feng10,huang12}.
It is shown that the magnetic field penetration depth shows a crossover
from the linear temperature dependence at low temperatures to a nonlinear
one at the extremely low temperatures. In particular, in analogy to the
domelike shape of the doping dependent SC transition temperature, the
maximal superfluid density occurs around the critical doping, and then
decreases in both lower doped and higher doped regimes. These are
consistent with the present results of the Raman response in the $n$-doped
cuprate superconductors and our recent results \cite{geng10} of the Raman
response in the $p$-doped cuprate superconductors, and all these studies
are important to confirm the nature of the SC phase of both $p$-doped
and $n$-doped cuprate superconductors as a d-wave BCS-like SC state
within the kinetic energy driven SC mechanism.

\acknowledgments

The authors would like to thank Dr. Zheyu Huang for helpful
discussions. This work was supported by the National Natural Science
Foundation of China under Grant No. 11074023, and the funds from the
Ministry of Science and Technology of China under Grant No.
2011CB921700.

\end{document}